\documentclass{llncs}
\usepackage{makeidx}    % allows for indexgeneration
\usepackage{listings}   % for source code listings
\usepackage{epsfig}
\begin{document}
\frontmatter          % for the preliminaries
\mainmatter              % start of the contributions
\title{Dynamic Web File Format Transformations with Grace}
\titlerunning{Dynamic Web File Format Transformations with Grace}  % abbreviated title (for running head)
\author{Daniel S. Swaney \and Frank McCown \and Michael L. Nelson}
\authorrunning{D. S. Swaney, F. McCown, M. L. Nelson}   % abbreviated author list (for running head)
%
%%%% modified list of authors for the TOC (add the affiliations)
\tocauthor{Daniel S. Swaney (Old Dominion University),
Frank McCown (Old Dominion University),
Michael L. Nelson (Old Dominion University)}
\institute{Old Dominion University\\
Computer Science Department\\
Norfolk, VA  23529 USA\\
\email{$\{$dswaney,fmccown,mln$\}$@cs.odu.edu}
}

\maketitle              % typeset the title of the contribution

\begin{abstract}
Web accessible content stored in obscure, unpopular or obsolete formats
represents a significant problem for digital preservation.  The file
formats that encode web content represent the implicit and explicit
choices of web site maintainers at a particular point in time.  Older
file formats that have fallen out of favor are obviously a problem, but
so are new file formats that have not yet been fully supported by browsers.
Often browsers use plug-in software for displaying old and new formats, but
plug-ins can be difficult to find, install and replicate across all
environments that one may use.  We introduce Grace, an http proxy server
that transparently converts browser-incompatible and obsolete web content
into web content that a browser is able to display without the use of plug-ins.
Grace is configurable on a per user basis and can be expanded to provide an
array of conversion services. We illustrate how the Grace prototype transforms
several image formats (XBM, PNG with various alpha channels, and JPEG 2000) so
they are viewable in Internet Explorer.
\end{abstract}
\section{Introduction}
Data formats for Web-accessible digital content are continually changing.
Digital contents that are stored in older or unpopular formats are increasingly
in danger of becoming inaccessible to modern web browsers. For example, the
XBM image format was properly displayed by Microsoft Internet Explorer (IE)
version 4, but IE version 6 is unable to display this format. Older formats
like early pkzip, PostScript, and PICT images are no longer accessible or
viewable without locating and installing old or special software. Newer digital
formats are also likely to be inaccessible through a web browser until wide-spread
adoption of the format forces browser makers to support the new format natively.
Although the Portable Network Graphic (PNG) format has been around since 1997,
its format is still not completely supported by IE 6.

Often it is left to the end user to install plug-ins, software that must be
downloaded and installed separately from the web browser, in order to view
files with older or unpopular file formats. Installation of this software
is not portable; it must be installed at each client browser for the file
format to be properly displayed. For example, the JPEG 2000 image format
requires the user to install a plug-in to view .jp2 image files on IE and
Netscape. Although a motivated user might install the software on their home
computer, they may not have administrative privileges to install the software
in a public computer lab. Plug-in software will often age along with the file
format it interprets leaving users unable to find up-to-date plug-ins for
modern web browsers.

We are in the process of building Grace\footnote{Named after the Grace
Brothers department store in the BBC Comedy ``Are you being served?''},
an http proxy that transparently
converts unsupported digital objects into formats that are supported by a
user's browser without the installation of any supporting software.  The
user can create a profile with the Grace system that allows the user to view
on-line content in the exact same manner from any web browser. Grace performs
format conversion dynamically by using a set of format translation rules
that can be customized and personalized by the user.

\begin{figure}
\begin{center}
\scalebox{1.1}{\includegraphics[width=6.9cm,clip,viewport=0 0 260
100]{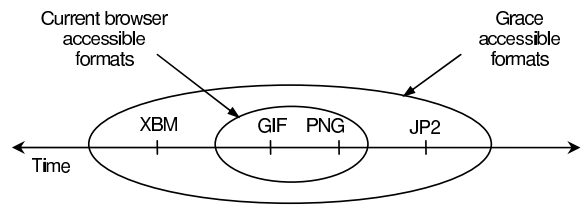}}
%\vspace{3.9cm}
\caption{Expanding time-line of browser-accessible formats using
Grace} \label{fig:grace_timeline}
\end{center}
\end{figure}

By using Grace, the accessibility of obsolete file formats like XBM
can be stretched over a longer period of time than is currently
possible with modern web browsers. New formats like JPG 2000 can be
viewed today. Figure \ref{fig:grace_timeline} illustrates how Grace
is able to expand the temporal bubble in which older and newer
formats are accessible in modern web browsers.

\begin{figure}
\begin{center}
\scalebox{0.5}{\includegraphics{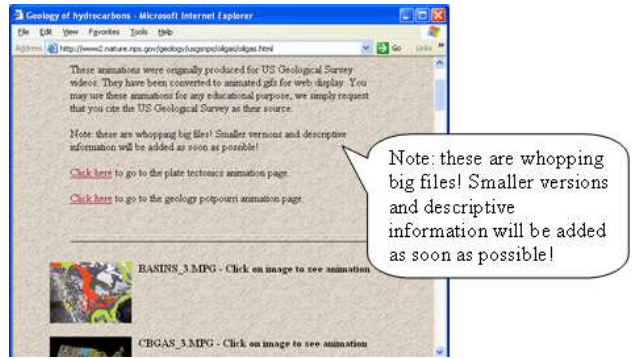}}
%\vspace{3.9cm}
\caption{``Whopping big files'' from the US Geological Survey, last modified on 2000-03-20}
\label{fig:usgs_whopping}
\end{center}
\end{figure}

Grace not only makes accessing web content easier for the end user,
it also supports a digital format preservation strategy which
relieves the web site operator from the burden of migrating web site
content that is stored using obsolete data formats. For example,
consider the on-line collection of MPEGs about the Geology of
Hydrocarbons\footnote{http://www2.nature.nps.gov/geology/usgsnps/oilgas/oilgas.html}.
There are 28 videos, each approximately 3 MB. Note the admonition on
the web page (Fig. \ref{fig:usgs_whopping}) promising new versions
of the videos. At the time of this writing, the page was last
modified on 2000-03-20. Other formats (e.g., .mov, .wmx, etc.) are
still not available, nor are the advanced video interfaces developed
by the Open Video Project \cite{march:geis} (e.g., storyboards, fast
forwards, etc.). It is unreasonable to expect the National Park
Service to pay for the ever-increasing storage demands for new
formats or pay for continued programming development for automatic
conversion. This page represents a valuable resource for education,
but there is currently no administrative or economic model to update
and upgrade the resources to take advantage of the latest formats
and interfaces.

In this paper we describe how the Grace system can convert a variety
of image formats into user-preferred formats.  Grace runs as a proxy
server and intercepts all web requests by the user. If an image is
requested that matches the user's translation rules, Grace uses
ImageMagick \cite{imagemagick}, an image format converter, to
convert the image transparently and return the transformed image to
the user. Grace allows additional software to be added internally or
externally for additional format conversions.
\section{Related Work}
The problem of obsolete data formats has been widely recognized for many years.
Solutions have primarily fallen into one of two camps: format migration and
software/hardware emulation, with the former being the most commonly used approach.

The Typed Object Model (TOM) addresses the problems inherent in accessing
data stored in obsolete data formats \cite{ocker1}. TOM provides the
ability to explain a data format, interpret the format for proper data
extraction, and convert the data to some other format. The TOM Conversion
Service \cite{ocker2} was built to demonstrate
how file format conversions could be performed using a web interface. It
uses third-party software to convert files uploaded by users. The Format
Registry Demonstration (FRED) \cite{fred} uses TOM to demonstrate how a global digital
format registry \cite{gdfr} may be created which provides
a central repository for information on new and obsolete file formats. We
are considering using the TOM service for performing external format
transformations in Grace. A global format registry could additionally be
useful for accessing data on obsolete data formats.

The JSTOR/Harvard Object Validation Environment (JHOVE) project is an
attempt to provide a set of services that can be applied to a variety
of digital formats \cite{jhove}. Given a digital object, JHOVE can determine what
format the object is stored in, it can determine if the object conforms
to a particular data format, and it can provide significant characteristics
of the object (i.e. width, height, language, encoding, etc.). JHOVE could
be used by Grace for determining the attributes of a web object so that
it can be properly converted into another format.  For example, if a web
server returned a web object with a MIME type of `image/gif', it could be
useful for Grace to know which version of GIF the object is using so an
appropriate conversion tool could be used if necessary.

The LOCKSS (Lots of Copies Keep Stuff Safe) project is used by many
libraries for preserving and making accessible on-line content
obtained by permission from a publisher \cite{reich:rose}.
LOCKSS is a P2P system which crawls a publisher's website and stores
local copies for safekeeping.  A web proxy server intercepts requests
for these web pages, and if they are no longer available from the
publisher's web site, the proxy server is able to serve a local copy
without the user being aware of the exchange. In order to address the
possibility of obsolete web formats among the digital content it is
preserving, LOCKSS has implemented a proof-of-concept system which
converts obsolete image formats into newer ones transparently by the
web proxy server \cite{rose:lipk}.  The goal is to create converters
which can be preserved along with the web content already being stored
in the LOCKSS system. The Grace system is similar to LOCKSS in its
desire to serve to the user web content that is transparently converted
into other web formats. But Grace is focused on individualized
transformations for any web-accessible content. Grace additionally
provides accessibility to web content using a variety of web browsers
and platforms; the LOCKSS migration facility only works on LOCKSS holdings.
\section{Grace Format Conversion}\label{sec:grace_format_conv}
A user can use Grace by configuring their web browser to proxy all
http traffic to the Grace Translation Service. All subsequent web
accesses will be directed through the Grace system which will return
all requested content and translated content to the web browser.

\subsection{High Level Design}
\begin{figure}
\begin{center}
\scalebox{1.3}{\includegraphics[width=6.9cm,viewport=0 620 330
800]{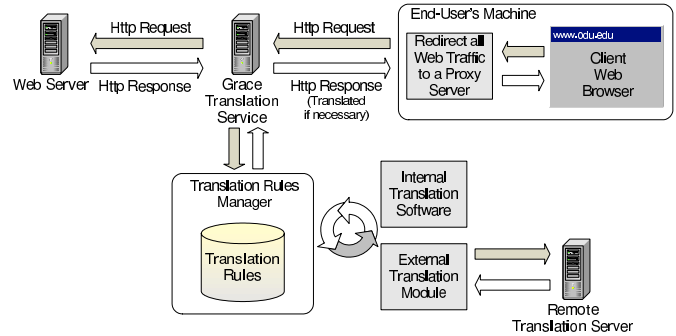}}
%\vspace{5.0cm}
\caption{High-Level Diagram of the Grace Translation Server}
\label{fig:high-level-diagram}
\end{center}
\end{figure}

Figure \ref{fig:high-level-diagram} shows how the Grace Translation Service transforms web
content between a client web browser and a web server. Each http
request is passed to the Grace Translation Service and then on to
the web server which responds with an http response back to the Grace
Translation Service.  The Translation Rules Manager uses the returned
response's MIME type to see if the user has setup any Transformation
Rules. Any matching rules cause the Internal or External Translation
Software to convert the content data.

ImageMagick \cite{imagemagick} is one software package used as Internal Translation Software.
ImageMagick can transform over 90 image formats including GIF, JPEG,
JPEG-2000, PNG, PDF, TIFF, and DPX. Additional third-party software can
be added as needed. An External Translation Module can be used for
transforming data formats using remote conversion services like TOM.

Once the content's data is transformed to a new MIME type, the
Translation Rules Manager will again search for another matching
rule for the new MIME type, and the process will repeat until no
more matching rules exist.  The transformed content is then returned
to the web browser as a new http response encoded with the new MIME type.

\subsection{Translation Rules}

The Grace Translation Server uses a set of XML encoded translation
rules for determining which types of web content to convert.
Each user has a profile which contains its own set of translation
rules.  Figure \ref{fig:trans_rules} shows two profiles. The `dswaney' profile has three
translation rules for converting JPG images into GIFs, XBM images
into PNGs, and GIF images into BMPs. The `mln' profile has a
translation rule for converting JPEG-2000 images into JPGs and GIFs
into PNGs. The rules are from the set of available transformations
show in Fig. \ref{fig:types-trans}. The user can use a web interface for
selecting the desired translation rules.

\begin{figure}

\lstset{language=XML} % set language to XML, can also be: c++ etc
\lstset{commentstyle=\textit,showstringspaces=false}   % more settings
\begin{lstlisting}[frame=trbl]{}    % enclose listing in box
<profile id="dswaney">
  <transform id="001" rule="JPG->GIF" />
  <transform id="002" rule="XBM->PNG" />
  <transform id="003" rule="GIF->BMP" />
</profile>
<profile id="mln">
  <transform id="001" rule="JP2->JPG" />
  <transform id="002" rule="GIF->PNG" />
</profile>

\end{lstlisting}
\caption{Two user profiles with transformation rules}
\label{fig:trans_rules}
\end{figure}

The Grace Transformation Server uses a list of XML encoded MIME
types and conversion software for performing the transformations.
As shown in Fig. \ref{fig:types-trans}, each transformation
defines the MIME type to be transformed
(mimetypesource tag), the target MIME type (mimetypetarget tag),
and the software to be used to perform the transformation (library tag).

\begin{figure}
\lstset{language=xml} % set language to XML, can also be: c++ etc
\lstset{commentstyle=\textit,showstringspaces=false}   % more settings
\begin{lstlisting}[frame=trbl]{}
<transformations>
  <transform id="JPG->GIF" description="Transform JPG->GIF">
    <mimetypesource>image/jpeg</mimetypesource>
    <mimetypetarget>image/gif</mimetypetarget>
    <library>TRImageMagick</library>
  </transform>

  <transform id="XBM->PNG" description="Transform XBM->PNG">
    <mimetypesource>image/x-xbitmap</mimetypesource>
    <mimetypetarget>image/png</mimetypetarget>
    <library>TRImageMagick</library>
  </transform>

  <transform id="JP2->JPG" description="Trans JPEG-2000->JPG">
    <mimetypesource>image/jp2</mimetypesource>
    <mimetypetarget>image/jpeg</mimetypetarget>
    <library>TRImageMagick</library>
  </transform>
</transformations>
\end{lstlisting}
\caption{Types of transformations that the Grace Transformation Server can perform}
\label{fig:types-trans}
\end{figure}

\section{Example Transformations}
We tested the Grace system on a variety of image formats using three tests:
\begin{enumerate}
\item Test formats that are commonly available by most web browsers.
\item Test the PNG format that is improperly displayed by a variety of web browsers.
\item Test the JPEG 2000 format which can only be displayed using a browser plug-in.
\end{enumerate}

We used Microsoft Internet Explorer (IE) 6 (with Service Pack 2)
since it is one of the most popular browsers used at the time of
writing. All the images transformed by Grace have a watermark at the
top-left part of the image to clearly show which images Grace has
transformed.  The watermark is shown only for testing purposes and
can be turned off during production use.
\subsection{Common Image Format Test}
\begin{figure}
\begin{center}
\includegraphics[width=10.0cm]{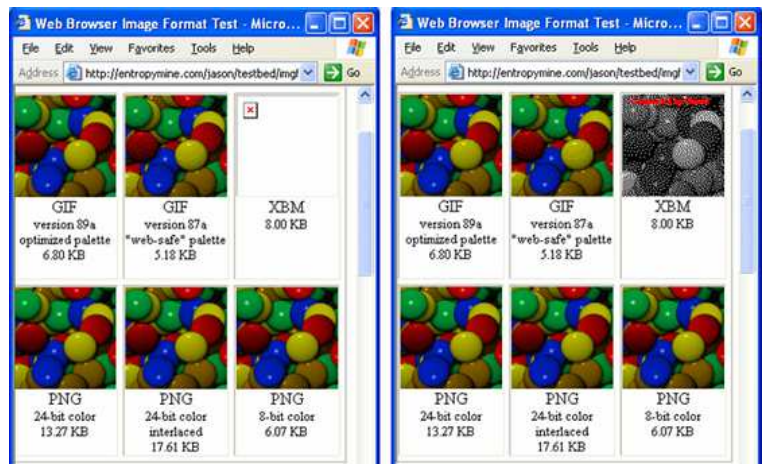}
\caption{IE is unable to display the XBM image (left). Grace transforms XBM into PNG image (right)}
\label{fig:xbm-png}
\end{center}
\end{figure}

The web page http://entropymine.com/jason/testbed/imgfmts/ contains
a variety of image formats (GIF, JPEG, PNG, and XBM) with various
color palettes and interlace options. As shown in Fig. \ref{fig:xbm-png}, IE 6
was able to display all the images except the XBM formatted image.
Although IE version 4 (circa 1997) was able to display the format,
it was abandoned by later versions of the browser. Using the Grace
system, the XBM format was converted into the PNG format which was
displayed properly in the browser.

\subsection{PNG Image Test}
Although the PNG image format has been incrementally supported
by a variety of browsers since its inception in 1997 \cite{roel}, there are
still many browsers that are unable to properly render the
various levels of alpha transparency that PNGs use.

\begin{figure}
\begin{center}
\includegraphics[width=5.0cm]{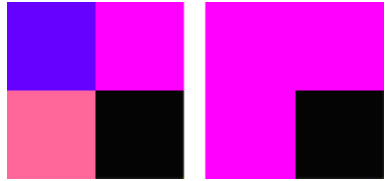}
\caption{Correct rendering of PNG (left). IE is unable to properly handle various PNG alpha thresholds (right)}
\label{fig:colored-pngs}
\end{center}
\end{figure}

The web page http://www.ecs.soton.ac.uk/\~{}njl98r/png-test/alpha/cmap.html
displays the various deficiencies that browsers have when displaying
PNG images. Figure \ref{fig:colored-pngs} shows an image on the left that is broken into
four smaller squares.  The three non-black squares all use various
levels of transparency. When viewed with IE, the image appears as
shown on the right side of Fig. \ref{fig:colored-pngs}.
IE is unable to display the transparent levels properly.

\begin{figure}
\begin{center}
\includegraphics[width=7.5cm]{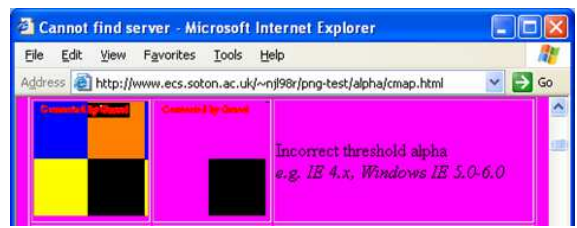}
\caption{Grace converts PNG into BMP}
\label{fig:grace-pngs}
\end{center}
\end{figure}

Grace was used to convert PNG images into BMP images,
a format that IE natively supports.  A portion of the
transformed web page is shown in Fig. \ref{fig:grace-pngs}. Although it
is not clear from a black-and-white rendering of this
paper, the image on the left side of Fig. \ref{fig:grace-pngs} contains
different colors because the transparent levels have now
been converted to a solid color by Grace. This is because
the BMP format is not capable of displaying transparency.
In this case, a lossy conversion results in image
attributes being lost.

\subsection{JPEG 2000 Image Test}

\begin{figure}
\begin{center}
\includegraphics[width=6.7cm]{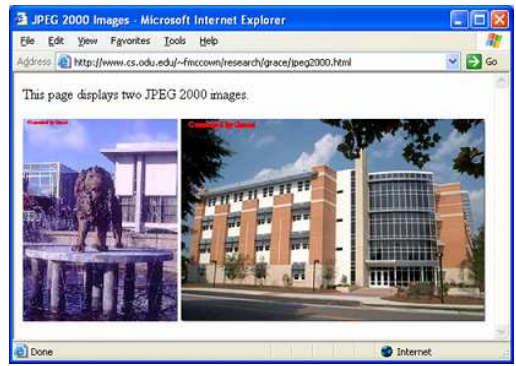}
\caption{IE is able to view JPEG 2000 images without a plug-in by using Grace}
\label{fig:jpeg2000}
\end{center}
\end{figure}

The JPEG 2000 image format is the successor to the JPEG
image format.  Being a new format, it is not supported
by the major web browsers except with use of a plug-in.
Grace is able to transform JPEG 2000 images into JPEG
images that all major web browsers can display. Figure \ref{fig:jpeg2000}
shows how IE can display a web page with two JPEG 2000
images that have been transformed by Grace into two JPEG
images.

\section{Future Work}
The Grace prototype shows how image formats can be
dynamically and transparently converted into other
formats. There is much work to be done in making Grace
perform other types of translations in a time efficient
and scalable manner.

\subsection{Additional Types of Translations}

The Grace prototype currently only translates image formats.
Future improvements to Grace will enhance its ability to
transform a wide range of data formats using TOM or other
conversion services. For example, a Microsoft PowerPoint
file could be converted into a PDF or into a series of web
pages that can be viewed in any browser.

Other Grace transformations may not focus on the MIME
type of a particular resource. Grace could be used to
transform an English version of a web page into Spanish
using a language translation application, or for stripping
out unwanted advertisements or extraneous information that
is not desired \cite{liu:ng}.  Grace could be used in transducing,
converting web content into formats that are more easily
viewable on small personal devices with limited viewing and
bandwidth capabilities \cite{trev:hilb}. These types of transformations
will provide more of a challenge than the digital format
conversion illustrated in this paper, but they would allow
Grace to be the single framework on which many user interface
transformation projects reside.

\subsection{Improve File Format Metadata}

Currently Grace only works with MIME types which do not provide a
level of granularity necessary for all translation rules. For
example, a user might want to create a translation rule that
converts all Microsoft Word documents earlier than version X into
version Y. The MIME type of Word documents is `application/msword'.
Although an optional parameter may also specify the version of the
Word document \cite{lind}, this information is not always given.
Therefore a service like JHOVE could be useful for determining the
version of a particular Word document in order for Grace to know if
the translation rules need to be applied to it.

\subsection{Faster Translation of Web Content}

Translation of web content at access time is potentially a slow
process and may introduce unacceptable delay times for users. Our
initial prototype did not suffer from any noticeable time delays,
but more rigorous performance testing will be necessary before Grace
is ready for wide-spread use. One method we will investigate for
improving performance is to cache proxy transformations. After a set
of transformations have been applied to a web page, other users
requesting the same page need not reconvert the same content.
Possibly some transformations can be performed before access:
content that is one click away could be transformed while the user
is viewing the current web page.

\subsection{Scalable Network of Grace Servers}

A single Grace server would be prone to a variety of scalability
issues. In order for Grace to be scalable, a network of cooperating
Grace servers could be used. Servers could communicate with each
other using Open Archives Initiative Protocol for Metadata
Harvesting (OAI-PMH) \cite{lagoze:sompel}, a protocol that is
suitable for any scenario that requires periodic updating of XML
encoded data. OAI-PMH could be used to transmit transformation
capabilities and functionality from one Grace server to another
using periodic harvests. This would allow new transformation
capabilities to be introduced anywhere within the network and allow
new servers to quickly learn of existing transformation
capabilities.

\subsection{Missing Web Content}

Often data stored in obsolete formats are overlooked
by web masters that are busy monitoring more
up-to-date material on their web sites. These
overlooked data files may be inadvertently deleted
from a web server. Web pages have been shown to be
extremely ephemeral as well which means content being
linked to today may be inaccessible a short time later.
When web content is missing, Grace could be configured
to transparently fetch the latest version of the missing
content from the Internet Archive \cite{archive}. This type of architecture
mimics what LOCKSS does with the content that it crawls and
stores locally.

\subsection{Translation Rules for Various Browser Capabilities}

Grace could be enhanced to perform automatic browser
capability checking. For example, Grace could check to
see if the browser is capable of displaying PDF documents.
If the required plug-in is not found, a translation rule
could automatically be applied which converts a requested
PDF into a set of web pages. This would make it easier for
users who are less technically savvy and may not know what
a PDF is or how to install plug-ins. This would also
relieve most users from having to manually create translation
rules.

\subsection{Transformation Metadata}

How will a user know if what they are viewing is the original
resource or a Grace-transformed resource?  If a conversion results
in a loss of information (as our PNG to BMP conversion did), how can
the user be informed?  How can metadata about each transformation be
made accessible?

There are many issues which must be addressed in future versions of
Grace to allow advanced users access to transformation metadata.
Grace is currently configured to place a watermark on top of images
that it converts, but the watermark can be distracting and cover
vital portions of the image, especially on small images. In order
for a user to know that something has been transformed by Grace, it
might be better to present a link to transformation metadata at the
top of the page or inline with transformed content. This solution
will unfortunately alter the look of the web page. Usability studies
will need to be performed to determine how best to present this
information to users without distracting or annoying them. Most
users may not find transformation metadata very useful for casual
browsing.

\section{Conclusions}
We have introduced Grace, an http proxy server that dynamically
and transparently converts web accessible content. Grace allows
a user to view web content in a consistent manner independent
of browser and plug-in software. Grace not only improves the
web surfing experience for the end-user, but it also frees the
content-provider from the sometimes costly and difficult process
of converting existing on-line content into newer/popular data
formats. Grace uses a set of translation rules for converting
the format of web content on a per user basis. We demonstrated
how Grace could be used to allow Internet Explorer version 6 to
display several image types (XBM, PNG with various alpha channels,
and JPEG 2000) that it is unable to display without the help of
plug-ins. Several improvements were offered for expanding the
types of translations possible, for improving the speed and
quality of translations, and for allowing cooperation among
Grace servers.


\end{document}